\documentclass[12pt]{article}
\usepackage{times}
\usepackage{geometry}
\usepackage[utf8]{inputenc}
\usepackage{enumitem,amssymb}
\usepackage{ragged2e}
\usepackage{indentfirst}
\usepackage[round,authoryear]{natbib}
\newlist{thematic}{itemize}{8}
\setlist[thematic]{label=$\square$}
\usepackage{pifont}
\usepackage{hyperref}
\begin{document}

\raggedright
\huge
The Nonbinary Fraction: Looking Towards the Future of Gender Equity in Astronomy \linebreak
\normalsize

\textit{A State of the Profession Consideration}\linebreak

Kaitlin C. Rasmussen\textsuperscript{1,2,*} (she/they), Erin Maier\textsuperscript{3}  (they/them), Beck E. Strauss\textsuperscript{4,**} (they/them), Meredith Durbin\textsuperscript{5} (they/them), Luc Riesbeck\textsuperscript{6}  (they/them), Aislynn Wallach\textsuperscript{5} (they/them), Vic Zamloot\textsuperscript{7} (they/them), Allison Erena\textsuperscript{8} (they/them)\newline

\textsuperscript{1}Department of Physics, University of Notre Dame, Notre Dame, IN 46556, USA\newline
\textsuperscript{2}Joint Institute for Nuclear Astrophysics and Center for the Evolution of the Elements, East Lansing, MI 48824, USA\newline
\textsuperscript{3}Department of Astronomy and Steward Observatory, University of Arizona, Tucson, AZ 85719, USA\newline
\textsuperscript{4}\textit{Independent researcher}\newline
\textsuperscript{5}Department of Astronomy, University of Washington, Seattle, WA 98195, USA\newline
\textsuperscript{6}Space Policy Institute, George Washington University, Washington, D.C. 20052, USA\newline
\textsuperscript{7}Irell and Manella Graduate School of Biological Sciences, The City of Hope, Duarte, CA 91010, USA\newline
\textsuperscript{8}Department of Astronomy, Smith College, Northampton, MA 01063, USA\newline

* Principal \newline
** Corresponding (Beck.E.Strauss@gmail.com)\linebreak

\textbf{Abstract:}

Gender equity is one of the biggest issues facing the field of astrophysics, and there is broad interest in addressing gender disparities within astronomy. Many studies of these topics have been performed by professional astronomers who are relatively unfamiliar with research in fields such as gender studies and sociology. As a result, they adopt a normative view of gender as a binary choice of `male' or `female', leaving astronomers whose genders do not fit within that model out of such research entirely. Reductive frameworks of gender and an overemphasis on quantification as an indicator of gendered phenomena are harmful to people of marginalized genders, especially those who live at the intersections of multiple axes of marginalization such as race, disability, and socioeconomic status. In order for the astronomy community to best serve its marginalized members as we move into the next decade, a new paradigm must be developed. This paper aims to address the future of gender equity in astronomy by recommending better survey practices and institutional policies based on a more complex approach to gender.

\newpage
\textbf{Co-signers}\newline
Rachael Amaro (she/her), Steward Observatory, University of Arizona\newline
Claudia Antolini (she/her), \textit{Independent researcher}\newline
Rachael L. Beaton (she/her), Princeton University and Carnegie Institution for Science\newline
Hayley Beltz (she/her), University of Michigan\newline
Ali M. Bramson (she/her), University of Arizona\newline
Jonathan Brande (he/him), NASA Goddard Space Flight Center/University of Maryland\newline
Kaley Brauer (she/her), Massachusetts Institute of Technology\newline
Harriet Brown (she/her), Science and Technology Facilities Council\newline
Katie Chamberlain (she/her), Steward Observatory, University of Arizona\newline
Jonathan Crass (he/him), University of Notre Dame\newline
Marion Cromb (they/them), University of Glasgow\newline
Mia de los Reyes (she/her), California Institute of Technology\newline
Adeene Denton (she/her), Purdue University\newline
Jeremy Dietrich (he/him), University of Arizona\newline
Serina Diniega (she/her), NASA Jet Propulsion Laboratory\newline 
Ewan Douglas (he/him), Steward Observatory, University of Arizona\newline
Courtney Dressing (she/her), University of California, Berkeley\newline
Tasha Dunn (she/her), Colby College\newline
Hannah Earnshaw (they/them), California Institute of Technology\newline
J.J. Eldridge (she/they), University of Auckland\newline
Rana Ezzedine (she/her), Massachusetts Institute of Technology and JINA Center for Evolution of the Elements\newline
Justin Filiberto (he/him), Lunar and Planetary Institute, Universities Space Research Association\newline
Anna Frebel (she/her) Massachusetts Institute of Technology\newline
Rachel L.S. Frisbie (she/her), Michigan State University\newline
Peter Garnavich (he/him), University of Notre Dame\newline
Juliane Gross (she/her), Rutgers University\newline
Hannah Holmberg (she/they), Smith College\newline
Deanna C. Hooper (she/her), RWTH Aachen University\newline
Macy Huston (she/they), Pennsylvania State University\newline
Raphael Erik Hviding (he/him), Steward Observatory, University of Arizona\newline
Alexander P. Ji (he/him), Carnegie Observatories\newline
James T. Keane (he/him), California Institute of Technology\newline
Zahra Khan (she/her), \textit{Independent researcher}\newline
Stanimir Letchev (he/him), University of Notre Dame \newline
Dan Marrone (he/him), Steward Observatory, University of Arizona\newline
Tia Martineau (she/her), University of New Hampshire\newline
Amelia Mangian (she/her), University of Illinois\newline
Rhiannon Mayne (she/her), Texas Christian University\newline
Bradley Meyers (he/him), Curtin University\newline
Moses Milazzo (he/him), Other Orb LLC\newline
Andrew Mizener (he/him), Macalester College\newline
Sarah E. Moran (she/her), Johns Hopkins University\newline
Henry Ngo (he/him), National Research Council of Canada\newline
Christine O'Donnell (she/her), University of Arizona\newline
Locke Patton (he/him), Harvard-Smithsonian Center for Astrophysics\newline
Cynthia Phillips (she/her), NASA Jet Propulsion Laboratory\newline
Chanda Prescod-Weinstein (she/they), University of New Hampshire\newline
Ragadeepika Pucha (she/her), University of Arizona\newline
Julie Rathbun (she/her), Planetary Science Institute\newline
Christina Richey (they/them), NASA Jet Propulsion Laboratory\newline
Jane Rigby (she/her), NASA Goddard Space Flight Center\newline
Edgard G. Rivera-Valent\'{i}n (they/them), Lunar and Planetary Institute, Universities Space Research Association\newline
Andy Rivkin (he/him), Applied Physics Laboratory, Johns Hopkins University\newline
C.W. Robertson III (he/him), Purdue University\newline
Ian U. Roederer (he/him), University of Michigan\newline
Gregory Rudnick (he/him), University of Kansas\newline
Catheryn Ryan (they/them), York University \newline
Julia E. Stawarz (she/her), Imperial College London\newline
H. F. Stevance (she/her), University of Auckland\newline
Kate Storey-Fisher (she/her), New York University\newline
Coleman A. Thomas (he/him), University of Notre Dame \newline
Meagan Thompson (she/her), University of Texas at Austin\newline
Sarah Tuttle (she/her), University of Washington, Seattle\newline
Aparna Venkatesa (she/her), University of San Francisco\newline
Monica Vidaurri (she/her), NASA Goddard Space Flight Center\newline
Lucianne Walkowicz (they/she), The Adler Planetarium\newline
MacKenzie Warren (he/him), North Carolina State University and Michigan State University\newline
Christopher Waters (he/him), Princeton University\newline
Henry Zovaro (he/him), Australian National University\newline

\newpage

\section{Introduction}

In the three decades since the first Women in Astronomy meeting in 1992 and the subsequent release of the Baltimore Charter for Women in Astronomy \citep{1993AAS...182.6501U}, there have been numerous studies, community efforts, and institutional initiatives addressing the status of women\footnote{In practice, the primary subjects and beneficiaries of such work have typically been limited to cisgender, white, heterosexual, abled women.} within astronomy. More recently, transgender and nonbinary identities have also gained recognition and inclusion in astronomy equity efforts. For instance, the American Astronomical Society (AAS) Committee for Sexual-Orientation and Gender Minorities in Astronomy (SGMA)\footnote{\url{https://sgma.aas.org/}} was established in 2012 and subsequently published the first and second editions of the LGBT+ Inclusivity Best Practices Guide \citep{2018arXiv180408406A}. The American Physical Society (APS) report ``LGBT Climate in Physics: Building an Inclusive Community'' \citep{APSLGBT} was released in 2016. Pronouns\footnote{In this context, ``pronouns" refers to the set of pronouns that should be used to refer to an individual person, such as ``they/them/theirs" or ``she/her/hers". See \url{https://www.mypronouns.org/} for more information.} are now optional to display on badges at many conferences, including meetings of the American Astronomical Society.\newline

However, many gender-related studies and equity initiatives have been led by professional astronomers with little to no background or training in gender studies or social sciences. As a result, they often treat gender using concepts and methods that range from reductive to actively harmful. In this white paper, we first summarize recent work on gender equity in the field of astronomy. We then discuss common pitfalls of such work, with a focus on its detrimental effects on the inclusion of nonbinary genders. Finally, we offer recommendations for studying gender dynamics and promoting gender inclusion in astronomy going forward.\footnote{For a more general consideration of the necessity of and avenues for collaboration between the space sciences and the social sciences, we direct readers to the companion white paper by \citet{Berea2019}.}\newline


Throughout this work, we use `nonbinary' as an umbrella term for all genders not represented by the categories of `male' or `female.' While this is the way most of our author list identifies, we would like to make clear that not everyone whose gender falls under this definition uses the term `nonbinary' to describe themselves, and that language surrounding gender identity is continually evolving and rarely universally agreed upon by those it purports to describe.\newline

\textit{N.B.:} The authors wish to make our own positionality clear: although we all identify as nonbinary, we are in no way representative of all nonbinary people and should not be misconstrued as speaking for all people who experience gendered marginalization. We also acknowledge that while we emphasize the importance of expertise that the social sciences bring to bear on this topic, we ourselves have collectively little formal training in such disciplines.

\section{The state of gender-related studies in astronomy}

A substantial number of recent studies have attempted to evaluate gender disparities within astronomy. Most of these publications were written by professional astronomers, and all were intended for and circulated among audiences comprised of astronomy and astrophysics researchers. While this literature review is far from comprehensive, the papers described here are broadly illustrative of typical concepts and methods that astronomers employ in studying gender-related phenomena within the field.\newline

Most commonly, these studies examine the impact of gender on career-related metrics. For example, \citet{2014PASP..126..923R}, \citet{2016Msngr.165....2P}, and \citet{2016arXiv161104795L} look at time allocation at major observatories by gender. \citet{2017NatAs...1E.141C} analyze citation rates of publications in major astronomy journals. \citet{2018arXiv181001511F} and follow-up study \citet{2019arXiv190308195P} both examine the length of time between the completion of the Ph.D. and hiring into a long-term position within or adjacent to astronomy.
\newline

Another topic of interest is social dynamics in professional settings. An ongoing series of publications, beginning with \citet{2014arXiv1403.3091D} and followed by \citet{2014A&G....55f.6.8P}, \citet{ 2016csss.confE.155T}, and \citet{2017NatAs...1E.153S}, assesses the likelihood of a person asking a question in a conference session as a function of the gender of the question-asker, the gender of the speaker, and the gender of the session chair. 
\newline

Two other noteworthy publications released this year concern methods intended to promote gender equity in astronomy. \citet{2019arXiv190503314H} describe an algorithm, Entrofy, which can be used in cases of cohort selection---i.e., to maximize diversity along committee-defined axes such as race, gender, career stage or seniority, skill set, and geographic origin when admitting students, awarding grants, choosing conference speakers, and so forth. \citet{oliveira2019} present an initiative implemented at the Space Telescope Science Institute to improve gender representation on institutional committees, and to track gender representation in activities that the committee organizes or contributes to, such as the selection of invited speakers for a conference.

\subsection{Problematic approaches to gender and their impacts}

We identify three major concerns common to most or all of the analyses presented in the above works: (1) the treatment of gender as observable through means other than self-identification; (2) categorization schemes with limited gender options; and most critically, (3) an over-reliance on quantitative methodology that is at best insufficient for understanding gendered phenomena in astronomy, and at worst epistemically violent towards people whose genders are poorly represented by these schema. Moreover, these studies typically demonstrate little engagement with the vast bodies of relevant existing work in such disciplines as gender studies and sociology, nor do they prioritize the testimony and participatory inclusion of people of marginalized genders.

\subsubsection{Gender as observable}
These studies, explicitly or implicitly, rely on gender information acquired by means other than participant self-identification. Most frequently, subjects are assigned a binary gender using first names; in some cases this is based upon the authors' own perceptions, while others make use of automated methods such as the Python package SexMachine\footnote{\url{https://pypi.org/project/SexMachine/}}, the GlobalNameData\footnote{\url{https://github.com/OpenGenderTracking/globalnamedata}} database, and the Gender API\footnote{\url{https://gender-api.com/}}. In cases where gender is indeterminate based on a name alone, some studies search for other public records, such as photos or articles including third-person gendered pronouns, in order to infer gender; most simply remove all data points with indeterminate names. Still other studies obtain gender data from volunteer data collectors' reports of subjects' (perceived) gender in real time.\newline

This treatment of gender as trivially discernible through names, physiology, or any other such external characteristics is unavoidably discriminatory\footnote{For an extensive discussion of flaws in the premises of algorithmic gender determination, and the myriad negative consequences transgender people experience from it, see \citet{Keyes2018}; while the critique focuses on automated gender recognition from images, many of their arguments are applicable more broadly to other external assessments of gender.}. For nonbinary people in particular, there is simply no acceptable outcome here: we are either misclassified into a binary gender, or considered uncategorizable and discarded. While this may sound trivial, experiences of misgendering and erasure have very real psychological and professional consequences for nonbinary, transgender and gender non-conforming individuals \citep{grant2011, McLemore2015, Davidson2016, Mizock2017, Thorpe2017, Cech2017Q, Cech2018}.\footnote{For example, the authors of this paper have been so aggrieved by the studies described here that we collectively assembled, organized, and wrote a decadal white paper about it.}

\subsubsection{Gender as discrete}

A closely related corollary to the above is that gender in these works is always treated as a set of discrete categories, which are implicitly presumed to be stable and coherent across populations, within individuals, and over time. The majority employ the male/female binary as a matter of course, and while the Entrofy algorithm presented in \citet{2019arXiv190503314H} does not require gender to be strictly binary, it requires that it be discret\emph{izable}. Specifically, the authors state that ``while we appreciate that gender is not a discrete concept, we advise against modeling gender as a continuous variable here because the proposed algorithmic framework requires an order relation over continuous values" \citep{2019arXiv190503314H}. Such frameworks inherently reduce members of a category to interchangeable data points, which loses much nuance and ultimately risks denying subjects authority over how their identity is represented.

\subsubsection{Gender as statistic}

Several of the works we have described include statements to the effect of: ``While we recognize that gender is not binary, we do not include nonbinary people in our analysis due to lack of statistical significance."\footnote{While we recognize that such disclaimers are well-intentioned, we do not count them in our assessments of said papers due to their lack of significance as anything other than empty gestures.} Statistical significance is here the determining factor in who gets to be accounted for---who \emph{counts}. In this way, reducing the work of inclusion to that which is quantifiable and measurable produces simplistic results that fail to describe the deep nuance and complexity of gender and the experiences of people navigating it within astronomy.\newline

Additionally, such complexities cannot be properly understood without also considering race, dis/ability, and other axes of marginalization \citep{Combahee1986, Crenshaw1989, Crenshaw1991,PrescodWeinstein2016,2017NatAs...1E.145P}. Even the term ``gender" has several shades of meaning--it can refer to the way someone is perceived, the way they are treated, and/or the way they see themselves. These concepts are all heavily influenced by other aspects of identity and social context, and are far from invariant across cultures \citep{Oyewumi1997, McLelland2005, chiang2012, kugle2014, besnier2014, bhaduri2015}. Serious conversations about gender equity must reckon with gender \emph{complexity}.

\subsection{Reinventing the wheel}
Gender is not a new area of study and gender inclusion is not a new problem. There have been efforts to make improvements in other disciplines for decades, and gender studies has existed as a field for at least the last century. Rather than deferring to established and well-researched understandings of gender and its individual and structural manifestations, many astronomers have chosen to attempt to solve problems of gender inclusion without consulting those whose fields have long been focused on this work. The hubris of this approach threatens to undermine the urgency of the problem at hand, requiring many astronomers to relearn lessons and rediscover concepts that are commonplace in other fields and, in the process, risk causing harm to those they intended to help. It further acts to frame the problem as one whose solution does not require the expertise of marginalized people themselves. 

\section{Recommendations}

Ultimately, we recognize that however flawed their approaches, the studies and other initiatives that we have discussed have come from a genuine desire to effect positive change within our field. Therefore, for the future, we make these recommendations.

\subsection{Methodological choices}

Approaches to gender, qualitative or quantitative, cannot come without a deep awareness of how complex, and contextual, gender is as a phenomenon. Prior to any gender research, responsible researchers \emph{must} begin by thoroughly grounding the precise definition of `gender' they seek to employ or investigate, the alignment of their data collection and analysis methods with said meaning, and the parameters of said data's utility. \newline

Despite our critiques of quantitative methods, we are not proposing to do away with them entirely; rather, what we are proposing is a community-wide reconsideration of the epistemic authority of such methods in matters of marginalization. Qualitative data and methods such as ethnographic description and participant testimony must be understood as valuable (and funded accordingly), rather than dismissed as `anecdotal' or `subjective'. Many examples of ethnographic work examining gender and race already exist in astronomy, including but not limited to \citet{2018PRPER..14a0146G, 2017JGRE..122.1610C, 2013AIPC.1513..222K, 2013AIPC.1517...33H}, and \citet{2011AAS...21743106G}. 

\subsection{Collecting and reporting gender data}

We emphatically discourage gathering gender data through any means other than voluntary self-identification, and doubly discourage the use of automated gender classification methods \citep{Keyes2018}. We encourage the Decadal Survey Committee to recommend that journals and funding agencies prioritize gender equity initiatives that use the best practices described in this paper and/or are conducted in collaboration with social scientists.\newline

For the collection of demographic data, we recommend use of the template provided by the Open Demographics initiative\footnote{\url{http://nikkistevens.com/open-demographics/questions/gender.html}}, which includes survey questions on four facets of gender: cisgender/transgender status, gender conformity, intersex status, and gender identity. Respondents should have the options to specify that they prefer not to disclose this information, or to refuse to answer the question entirely: for some, even a refusal to disclose may be perceived as too revealing.\newline

In cases such as allotting observing time where anonymity is desired, anonymity should be preserved throughout the assignment and the demographic information used only after its completion as a check on the system. In cases where anonymity is not sought, such as conference speaking slots and job applications, use of the Open Demographics template will prevent harmful assumptions from being made. In this scenario, however, we caution that if gender identity is intentionally withheld, that decision should be respected.\newline

In all cases we \textit{strongly} encourage the employment, or at least compensated consultation, of trained social scientists when studying marginalized people in astronomy. We therefore call for the creation and support of dedicated funding sources for interdisciplinary research in space science, with a focus on enabling collaboration by providing material support to experts in other fields (e.g., grants incorporating a Co-I or Collaborator status).\newline

For further reading on gender-inclusive data collection and survey design, we recommend \citet{geniuss2014, Labuski2015, Westbrook2015, Doan2016, Magliozzi2016, Broussard2018, Glick2018} and references therein.

\subsection{Privacy considerations}
Gender data should not be shared outside the context for which it was collected; for instance, if researchers wish to analyze demographic data for a particular conference, they should collaborate with conference organizers to institute collection of the data in a responsible, transparent way. Explicit and rigorous precautions should be taken to protect participant privacy when collecting and presenting any kind of identifying information, and said precautions should be made clear to participants beforehand (e.g., by specifying what offices may handle personally identifiable information and publicly posting policies on these practices).\newline 

Astronomy is a fairly small, well-connected community, and marginalized individuals, particularly those who are multiply marginalized, can sometimes be uniquely identified through demographic data even in the absence of other identifying information. Thus, when \emph{reporting} gender statistics, it is acceptable to group all nonbinary identities as one class (e.g., ``We find that X percent of women, Y percent of men, and Z percent of nonbinary individuals..."). This practice should be avoided in survey design and data collection, but in cases where anonymity in reported results is a concern, grouping nonbinary respondents may help to preserve privacy.

\subsection{Institutional and educational policies and practices}

A full consideration of gender equity in the context of institutional reform is beyond the scope of this paper. Our recommendation is this: gender equity requires the adoption of a more complex model of gender than has historically been employed by equity initiatives. Many organizations and committees currently focus only on women in astronomy. Thus, we recommend that this focus be shifted to people of marginalized gender in astronomy. We recognize that this is a major, fundamental change to the status quo, but as we move into the next decade of astronomy research, it is a change that must be made in order to support all members of the astronomy community.\footnote{Although \citet{Traxler2016} focus on physics education research, many of their approaches to this kind of systematic change can be applied in the context of astronomy research.}\newline

In order to facilitate such a systemic change, astronomers need to become familiar or at least acquainted with such disciplines as gender studies, critical race theory, ethics, and STSS\footnote{Science, Technology, and Society Studies}, while recognizing the limitations of our knowledge. We suggest that study in these areas be added to undergraduate and graduate curricula, as well as workforce development programs.\newline

We also recommend that telescope Time Allocation Committees move to an anonymous peer review model as recently done by the Hubble Space Telescope, and use the Open Demographics template described above if collecting demographic information from proposers.\newline

For additional guidelines on building inclusive physics and astronomy communities and institutions, we direct readers to the the Nashville Recommendations\footnote{\url{https://docs.google.com/document/d/1JipEb7xz7kAh8SH4wsG59CHEaAJSJTAWRfVA1MfYGM8/}} and the LGBT+ Inclusivity Best Practices Guide \citep{2018arXiv180408406A}.

\subsection{Final recommendations}

Our final, and perhaps most important recommendation, is to \emph{listen}. Look around your communities to see who the most marginalized, most vulnerable members are and make sure their voices are not just included but prioritized in conversations about equity and inclusion---that their needs and ideas are heard and valued.

\subsubsection{A note on consulting people of marginalized genders}

Marginalized people are frequently called upon to educate others about the conditions of their marginalization, typically without compensation \citep{Fricker2007, Berenstain2016}. We recommend that any astronomer wishing to educate themselves about gender first seek out resources independently to the best of their abilities. In particular, astronomers wishing to collect demographic information in surveys or for the purposes of cohort selection are encouraged to consult the works cited in this paper. However, we welcome readers to direct any remaining questions to the corresponding author of this paper.

\section{Cost estimates}

Historically, the true cost of studies like those critiqued here have been unknown numbers of students, postdocs, early career researchers, and other scientists who have found themselves alienated and excluded from the astronomy community.\footnote{We anticipate that any astronomer who chooses \textit{not} to conduct such a study will incur costs on the order of \$0.00.}  \newline

Hiring social scientists to do this work responsibly and covering the associated institutional costs will, of course, require additional funds \citep{DeptLabor2019}. \citet{Sell2017} offers a brief discussion of the financial costs associated with gender-inclusive survey design. We encourage interested astronomy researchers to anticipate and plan for such costs, and to seek out appropriate funding to collaborate with these experts.

\pagebreak

\bibliographystyle{apsr}
\bibliography{library}

\end{document}